\documentclass[11pt,twoside,smallextended]{article}
\usepackage{amsfonts,epsfig}
\usepackage{amsthm}
\usepackage{xy}
\input{xy}
\xyoption{all} \xyoption{rotate}

\theoremstyle{remark}

\newtheorem{definition}{Definition}[subsection]

\theoremstyle{plain}

\begin{document}

\title{Connectome graphs and maximum flow problems}
\author{
Peteris\ Daugulis\\ Department of Mathematics \\Daugavpils
University\\
Daugavpils, Latvia \\
peteris.daugulis@du.lv }

\maketitle

\begin{abstract} We propose to study maximum flow problems for connectome
graphs.  We suggest a few computational problems: finding vertex
pairs with maximal flow, finding new edges which would increase
the maximal flow. Initial computation results for some publicly
available connectome graphs are described.

\end{abstract}

\setcounter{tocdepth}{3} \tableofcontents

\section{Introduction} \label{intro-sec}

\subsection{The subject of study}

Connectome graphs are discrete mathematical models used for
modelling nervous systems on different scales, see \cite{K},
\cite{S1}, \cite{S2}. On the microscale level these graphs are
special cases of \sl cell graphs,\rm\ see \cite{B} for an example
of cell graph application in tumour tissue modelling. On mesoscale
and macroscale levels connectome graphs are essentially quotient
graphs of microscale cell graphs. In this paper we do not deal
with connectome scale and other modelling issues, we are
interested only in applications of graph-theoretic concepts and
algorithms. Connectome graph edges may be directed, undirected and
weighted (labelled). We consider only the directed graph structure
of connectomes, each edge is assigned a fixed weight. We use data
available at {http://www.openconnectomeproject.org}, \cite{O}, and
{connectome.pitgroup.org}, \cite{P}.

\subsection{Maximum flow and minimal cut problems}

We assume that the reader is familiar with basic definitions of
graph theory, see \cite{D}. We consider connectome graphs as being
directed loopless graphs.

We remind the basic facts about maximum flow and minimal cut
problems, see \cite{C}. Let $\mathcal{N}=(V,E,c)$ be a \sl
network\rm\ - $V$ is a finite set, $E\subseteq V\times V$,
$c:V\times V\rightarrow \mathbb{R}^{+}$ - the edge capacity
function such that $c(u,v)>0$ iff $(u,v)\in E$, $c(u,v)=c(v,y)$.
Let $s,t\in V$, a \sl $(s,t)$-flow\rm\ from $s$ to $t$
(single-source and single-target flow) is a function
$f_{s,t}:E\rightarrow \mathbb{R}^{+}\cup \{0\}$ satisfying $2$
conditions:

\begin{enumerate}

\item[1)] \sl capacity constraints \rm\ : - $0\le f_{s,t}(u,v)\le
c(u,v)$,

\item[2)]\sl\ conservation constraints\rm\ : - if $v\ne s$ and
$v\ne t$ then $$\sum_{u\in \Gamma_{-}(v)}f_{s,t}(u,v)=\sum_{w\in
\Gamma_{+}(v)}f_{s,t}(v,w).$$
\end{enumerate}

The \sl value of a flow \rm\ $f_{s,t}$ is defined as
$$\Phi(f_{s,t})=\sum_{u\in \Gamma_{+}(s)}f_{s,t}(s,u)=\sum_{v\in
\Gamma_{-}(t)}f_{s,t}(v,t).$$ Denote the set of all $(s,t)$-flows
by $F(\mathcal{N},s,t)$. Given a network $\mathcal{N}$ and two
vertices $s,t$ the \sl maximum flow problem\rm\ is concerned with
finding $$M(\mathcal{N},s,t)=\max\limits_{f_{s,t}\in
F(\mathcal{N},s,t)}\Phi(f_{s,t}).$$ If $s$ and $t$ are in
different weakly connected components of $\mathcal{N}$, then
$M(\mathcal{N},s,t)=0$. Computational complexity of algorithms
implemented in computer algebra systems for solving the maximum
flow problem is polynomial (at most cubic) in $|V|$ and $|E|$.

Let ${S,T}$ be a partition of $V$. An edge $(u,v)$ is a \sl
$(S,T)$-forward edge\rm\ if $u\in S$ and $v\in T$. An edge $(u,v)$
is a \sl $(S,T)$-backward edge\rm\ if $u\in T$ and $v\in S$.

$C\subseteq E$ is a \sl $(s,t)$-cut\rm\ if there are $S,T$ such
that $C$ is a union of $(S,T)$-forward and $(S,T)$-backward edges.
The capacity of a $(s,t)$-cut is the sum of capacities of all
forward edges. A classical result is the Max-Flow Min-Cut Theorem
due to L.Ford and D.Fulkerson - the maximal value of a
$(s,t)$-flow is equal to the minimal capacity of a $(s,t)$-cut.

Apart from single-source and single-target flows one can also
study flows having multiple sources and targets.

\subsection{Main objectives and steps of our work}\label{12}

We propose to consider directed connectome graphs as networks and
study maximum flow problems of these networks. Although maximum
flow problems originated in the transportation network science, it
makes sense to consider this problem in biological
discrete-mathematical models, e.g. in blood flow or signal flow
modelling. In the case of connectomes we could hypothesize that
directed connectome edges conduct flows of some sygnals or other
activities. We will consider unweighted directed connectome graphs
and assume that each directed edge has capacity $1$. In such a
case a maximum $(s,t)$-flow can be interpreted in terms of basic
graph-theoretic concepts - it is a maximal set of directed
edge-disjoint $(s,t)$-paths. Its possible relevance in modelling
of nervous systems can be based on an assumption that a connection
(edge) of a connectome graph can not be involved in more than one
sygnalling/activity process at a time.

We now describe some research direction involving maximum flow
problems.

\subsubsection{$(s,t)$-vertex pairs of maximal flow}

Given a network $\mathcal{N}$ we can pose the problem of finding
source-target vertex pairs admitting a flow with the maximal
possible flow value, we call such vertex pairs \sl extremal
pairs.\rm\ Assuming that flows have some biological
interpretation, extremal pairs would show optimal flow directions.
We can also compare maximal and average flow values.

\subsubsection{Restricted vertex pairs}

Given a source vertex $s$ and a target vertex $t$ we have that
$$M(\mathcal{N},s,t)\le \min(\deg^{+}(s),\deg^{-}(t)).$$

Vertex pairs with strict inequality have, in some sense, redundant
outgoing and ingoing edges. We can start looking for such vertex
pairs and interpret them in terms of the connectome graph
structure.

\subsubsection{Adding edges which increase maximal flow}

The ultimate goal of brain studies is to improve and develop the
human brain, make it more efficient and complex. Any advance in
mathematical modelling of the brain must be screened with respect
to this goal. Studying maximal flows in connectome graphs and
assuming that flows are important we can ask the following initial
questions: 1) what new (extra) edges would increase the maximal
flow? which edges are redundant ? 2) how can we add an extra
vertex and some new edges to maximize the maximal flow? which
vertices are redundant? In the simplest model when all edges have
capacity $1$, an extra edge can increase the maximal flow by at
most $1$ (an edge of capacity $c$ increases the maximal flow by at
most $c$). Thus, concerning the extra edge problem, we can only
look for nonadjacent vertex pairs $(a,b)$ such that the extra edge
$(a,b)$ would increase the flow by $1$.

\subsubsection{Summary}

The main objective of our work is to advertise the suggestion to
study maximum network problems for connectome graphs, present some
initial computations:

\begin{enumerate}

\item[1)] find maximal flows of connectomes and special vertex
pairs,

\item[2)] find vertex pairs which would define extra edges
increasing the maximal flow.

\end{enumerate}

\section{Main results}

\subsection{Definintions - vertex pairs with special properties}

$\mathcal{N}=(V,E,c)$ a network.

\begin{definition} An (ordered) pair $(a,b)$ is called an \sl extremal
$\mathcal{N}$-pair\rm\ if
$$M(\mathcal{N},a,b)=\max\limits_{s,t\in
V}M(\mathcal{N},s,t).$$

\end{definition}

%

\begin{definition} $\max\limits_{s,t\in V}M(\mathcal{N},s,t)$ is
called the \sl maximum flow $M(\mathcal{N})$\rm\ of $\mathcal{N}$.

\end{definition} The \sl\ average flow\rm\ of a network $\mathcal{N}$
is defined as $\frac{M(\mathcal{N})}{|P|}$, where
$P=\{(s,t)|s,t\in V, s\ne t\}$.

\begin{definition}

\end{definition}

\begin{definition} An (ordered) pair $(a,b)$ is called an \sl
restricted $\mathcal{N}$-pair with difference $d$\rm\ if
$$M(\mathcal{N},a,b)<\min(\deg^{+}(a),\deg^{-}(b))$$  and
$d=M(\mathcal{N},a,b)-\min(\deg^{+}(a),\deg^{-}(b))$.

\end{definition}

\begin{definition} An (ordered) pair $(a,b)$ is called a \sl
flow-increasing $\mathcal{N}$-vertex pair, if $(a,b)\not\in E$ and
$$M(\mathcal{N}+(a,b))>M(\mathcal{N}).$$

\end{definition}

\subsection{Some examples}

In this subsection we describe some of our computational results
related to (single-source and single target) flows, extremal and
flow-increasing vertex pairs in connectome graphs. Some numerical
answers are rounded. For connectome graphs with number of vertices
not exceeding $2000$ the maximum flow, extremal pairs can be found
on a standard laptop computer. The flow-increasing edge problem is
computationally more time consuming, we have only solved it for
connectomes of at most $50$ vertices.

Considering a connectome network $(\mathcal{N},V,c)$ we assume
that each edge has capacity $1$.

\subsubsection{Cat}

Filename - Mixed.species\_brain\_1.graphml, available at

{http://www.openconnectomeproject.org}.

Graph description - strongly connected graph with $65$ vertices
and $1139$ edges, underlying undirected graph has vertex
connectivity $6$, diameter $3$, radius $2$, center has $23$
vertices, minimal degree $3$, maximal degree $45$.

Maximal flow is $40$, average flow $\sim 12$. There is one
extremal vertex pair - $(53,59)$ (file vertex numbering
preserved).

\subsubsection{Worm}

There are $3$ connectomes for C.elegans and P.pacificus available
at

{http://www.openconnectomeproject.org}. Connectomes are not
strongly connected. Maximal flow list - $57,9,9$. Average flow
list - $6.5,0.5,0.5$. Number of extremal vertex pairs - $5,3,2$.
Many extremal pairs have one common vertex.

Most flow-increasing vertex pairs are of form $(n,c)$ for a fixed
$c$.

%

\subsubsection{Macaque}

There are $4$ connectomes for Rhesus macaque available at

{http://www.openconnectomeproject.org}. Maximal flow list -
$11,28,29,69$. Average flow list - $1,1,9,9$. Number of extremal
vertex pairs from $1$ to several hundreds. Many extremal pairs
have one common vertex.

\subsubsection{Rat}

There are $3$ connectomes for Rattus norvegicus available at

{http://www.openconnectomeproject.org}. Maximal flow list -
$472,493,496$. Average flow list - $25,20,17$. Number of extremal
vertex pairs - $2,1,6$. Many extremal pairs have one common
vertex.

\subsubsection{Mouse}

There are $4$ connectomes for mouse available at

{http://www.openconnectomeproject.org}, the maximal graph has
$1123$ vertices. Maximal flow list - $2,2,140, 540$. Average flow
list - $0.001,0.1,80,0.01$. Number of extremal vertex pairs - from
$1$ to $6$. Many extremal pairs have one common vertex.

For one small connectome ($29$ vertices) most flow-increasing
vetex pairs are of form $(n,c)$, for a fixed $c$.

\subsubsection{Fly}

Filename - drosophila\_medulla\_1.graphml, available at

{http://www.openconnectomeproject.org}. Graph description - $1781$
vertices and $9735$ edges, $996$ strongly connected components -
one with $785$ vertices, one with $2$ vertices, the other
components trivial, underlying undirected graph is disconnected,
has $6$ connectivity components (one big component - $1770$
vertices, connectivity $1$, $265$ cutvertices, diameter $6$,
radius $3$, center has $1$ vertex), minimal degree $1$, maximal
degree $927$.

Maximal flow - at least $100$. Average flow $\sim 10^{-6}$.

\subsubsection{Human}

Human connectome graphs available at
{http://www.openconnectomeproject. org} having about $800000$
vertices can not be processed in reasonable time using our
computing resources. An averaged human connectome graph example of
size which can be processed by a laptop ($1015$ vertices) is
available at {connectome.pitgroup.org}.

Graph description - $1015$ vertices and $8507$ edges, $1015$
trivial strongly connected components (no nontrivial ones),
underlying undirected graph is disconnected, has $84$ connectivity
components (one big component - $932$ vertices), minimal degree
$0$, maximal degree $204$.

Maximal flow - about $50$. Average flow $<10^{-4}$. Number of
extremal vertex pairs is $1$. Number of restricted vertex pairs is
about $30000$, all restricted pairs have difference $d=-1$.

\subsection{Conclusion}
We introduce study of maximal flow problems of connectome graphs
and present some initial computation results. For graphs having
about $2000$ vertices it is possible to compute maximal flows and
find extremal vertex pairs in several hours on a typical $2010$s
laptop computer. The problem of flow-increasing edges is more time
consuming, graphs of about $50$ vertices take several hours.

Some of our observations:
\begin{enumerate}

\item[1)] the maximal flow is significantly (by a factor of at
least $10$) larger than the average flow, the number of extremal
$(s,t)$-vertex pairs is small ($1$ in some cases);

\item[2)] for all graphs there are restricted vertex pairs;

\item[3)] for all graphs there are flow-increasing pairs, in many
cases one vertex of flow-increasing pairs is fixed or belongs to a
small subset of vertices.

\end{enumerate}

Further work can be done in the following directions:
\begin{enumerate}

\item[1)] interpret the computational results in biological and
modelling terms;

\item[2)] interpret the maximum flow and minimum cut problems for
models on different scales, consider single and multiple
source/target cases;

\item[3)] relate the known properties of "rich club" (various
centrality invariants) in terms of the network flow problem.

\item[4)] relate the structure of strongly connected components
and the Hertz graph in terms of maximal flow.

\end{enumerate}


\end{document}